\documentclass[doublecol]{epl2} 

\usepackage{graphicx}

\title{Experimentally observed evolution between dynamic patterns and intrinsic localized modes in a driven nonlinear electrical cyclic lattice}
\shorttitle{Experimentally observed evolution between dynamic patterns and ILM} 

\author{S. Shige\inst{1} \and K. Miyasaka\inst{1} \and W. Shi\inst{1} \and Y. Soga\inst{1} \and M. Sato\inst{1} \and A. J. Sievers\inst{2}}
\shortauthor{S. Shige \etal}

\institute{                    
  \inst{1} {Graduate School of Natural Science and Technology, Kanazawa University -- Kanazawa, Ishikawa 920-1192, Japan}
  \inst{2} {Laboratory of Atomic and Solid State Physics, Cornell University - Ithaca, NY 14853-2501, USA}
}

\abstract{
Locked intrinsic localized modes (ILMs) and large amplitude lattice spatial modes (LSMs) have been experimentally measured for a driven 1-D nonlinear cyclic electric transmission line, where the nonlinear element is a saturable capacitor. Depending on the number of cells and electrical lattice damping a LSM of fixed shape can be tuned across the modal spectrum. Interestingly, by tuning the driver frequency away from this spectrum an LSM can be continuously converted into ILMs and visa versa. The differences in pattern formation between simulations and experimental findings are due to a low concentration of impurities. Through this novel nonlinear excitation and switching channel in cyclic lattices either energy balanced or unbalanced LSMs and ILMs may occur. Because of the general nature of these dynamical results for nonintegrable lattices applications are to be expected. The ultimate stability of driven aero machinery containing nonlinear periodic structures may be one example.}

\begin{document}

\maketitle

\section{Introduction}
Studies of nonlinear lumped element electrical transmission lines have a long history starting with the observations of solitons in integrable lattices\cite{1,2} and such electric nonlinear lattices have continued to play an important role in the study of nonintegrable lattices\cite{3,4,5}. The idea that vibrational energy produced by nonlinearity in a discrete lattice could appear as intrinsic localized modes (ILMs) above the phonon spectrum\cite{6,7} now has a firm footing\cite{8,9,10,11,12,13,14,15,16}. [Such excitations are also called lattice solitons\cite{17} or discrete breathers (DB)\cite{16}.] A variety of experiments followed the initial theoretical developments with studies of macroscopic mechanical and electrical nonlinear transmission lines leading the way\cite{3,18,19,20}. For lumped element transmission lines there is a well known translation between inertial mass/spring force constants and electrical inductance/capacitance\cite{21}. Some of the experimental and theoretical works on nonlinear lattices have been summarized in three reviews\cite{22,23,24}. So far two directions have evolved: (1) to discover the properties of highly-localized ILMs and to assess their importance for real nonintegrable lattices and (2) to develop the general behavior of discrete nonlinear equations, with emphasis on identifying soliton-like behavior in lattices.

In this report the coupling between driven ILMs and pattern formation is explored by measuring the response of a 1-D nonlinear electrical transmission line. Although experimentally measured autoresonant (AR) ILMs are well known\cite{25,26} our observations of the associated tunable AR lattice spatial modes (LSMs) predicted by Burlakov\cite{14} are not. These latter modes are an example of modulational-instability mediated patterns in damped driven lattices. Here it is demonstrated that by tuning the driver frequency one can continuously shift the dynamical system from one nonlinear state to the other so that a cyclic AR LSM within the modal spectrum can become AR ILMs outside of that frequency region and visa versa. The energy asymmetries of the resulting nonlinear patterns are found to depend on a low concentration of impurities, the number of nonlinear units in the ring and the structure-damping factor.

\section{Process of LSM pattern generation}
\begin{figure}
\includegraphics{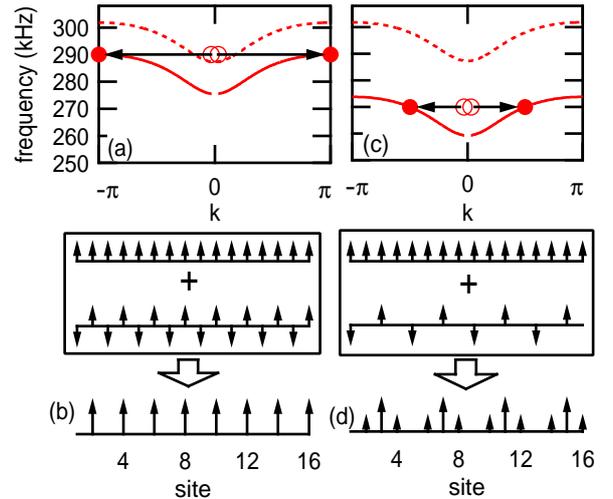}
\caption{\label{fig:figx}(a) LSM generation schematic in $k$-space. Dashed: linear dispersion curve without excitation. Solid: shifted dispersion curve due to large amplitude of the $k=0$ mode with uniform driver frequency represented by open circles. The $k=0$ uniform mode still resonates with the driver because of its very broad linewidth (low $Q$ system). Via 4-wave mixing two $k=0$ modes give rise to two zone boundary modes. (b) Instantaneous spatial pattern for the case in panel (a). $k=0$ uniform mode (top) and $k =  \pm \pi $  standing wave pattern (middle) vibrating at the same frequency. Total vector (bottom) is the eight peaked LSM. (c) Slightly lower driver frequency case than panel (a). Shift of the dispersion curve (solid) is now larger because the driver frequency is closer to the $k=0$ mode providing a larger resonance amplitude response. The four-wave scattering process now occurs at  $k =  \pm \pi /2$. (d) Instantaneous spatial pattern of the LSM for panel (c). Four peaks appear for the $N=16$ lattice. }
\end{figure}

There is value in first characterizing the four wave mixing process for LSM generation in a driven nonlinear lattice. Figure~\ref{fig:figx} is a schematic representation of the LSM generation process\cite{14} in $k$-space. The linear dispersion curve is represented by the dashed curve in frames (a) and (c). With the application of a uniform driver only the lowest point of the dispersion curve can be directly excited. Next imagine that the $Q$ of the system is small so that the resonance bandwidth extends over much of the dispersion curve. Finally include the property that the cyclic transmission line elements contain a soft nonlinearity. As the driver frequency approaches the dispersion curve from above the vibrational amplitude increases because of the broad resonance. With increasing $k=0$ amplitude, the dispersion curve shifts to lower frequencies due to the nonlinearity as shown by the solid curve in Fig.~\ref{fig:figx}(a), assuming all modes are at rest except the $k=0$ mode. At this point four-wave scattering $\left( {0,\Omega } \right) + \left( {0,\Omega } \right) \to \left( { - \pi ,\Omega } \right) + \left( { + \pi ,\Omega } \right)$ 
becomes possible because there exist normal nonlinear modes at $\left( {k,\omega } \right) = \left( {\pi ,\Omega } \right)$ and $\left( {-\pi ,\Omega } \right)$ in $k$-space. The modes at $\left( {k,\omega } \right) = \left( {\pi ,\Omega } \right)$ and $\left( {-\pi ,\Omega } \right)$ are excited parametrically. This represents the modulational instability of the $k=0$ mode. This excitation causes in real space the LSM by overlapping the $k=0$ uniform excitation and the $k =  \pm \pi $ standing wave vibrating at the same frequency as shown at one instant in Fig.~\ref{fig:figx}(b). For an even numbered lattice the number of amplitude peaks is $1/2$ the number of elements in the lattice. This explanation is admittedly schematic because once the $k =  \pm \pi $  modes acquire a large amplitude other normal modes (the rest of the dispersion curve) should be modified from the solid curve presented in Fig.~\ref{fig:figx}(a) by the nonlinear effects of  $k =  \pm \pi $ excitations. This modification may cause successive modulational instabilities, which would result in chaotic patterns in a high $Q$ lattice. For a low $Q$ lattice, considered in this paper, these additional instabilities can be suppressed resulting in a stable LSM.

	Burlakov\cite{14} actually studied the second modulational instability for the case of  $k =  \pm \pi /2$, which is represented by Fig.~\ref{fig:figx}(c). Although Figs.~\ref{fig:figx}(a) and (c) are schematic, they show at least the initial steps for generation of different LSM patterns. When the driver frequency is further reduced the nonlinear frequency shift of the dispersion curve grows because of the larger response of the system. The four wave scattering destination pair is switched to $k =  \pm \pi /2$  and a different LSM pattern is $1/4$ the number of elements in the lattice as shown at one instant of vibration in Fig.~\ref{fig:figx}(d).

\section{Experiments with a saturable capacitor}
\begin{figure}
\includegraphics{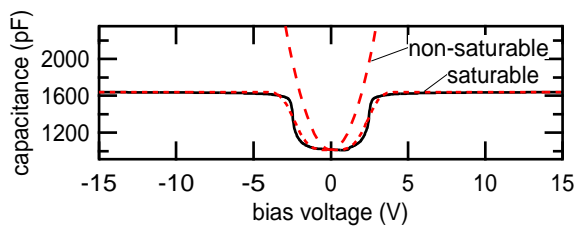}
\caption{\label{fig:fig1}Saturable and non-saturable capacitance versus bias voltage used in simulations. Capacitance as a function of the applied DC voltage. Saturable capacitor: solid curve, experimental measurement; dotted curve, fit used in simulations; nonsaturable capacitor, used in comparison simulations. }
\end{figure}

The 16 cell cyclic electrical transmission line circuit studied here is similar to that shown in Fig. 1 of Ref.~\cite{4} and described there in some detail. The same identification of the lumped elements is employed here for a cell with  linear inductor $L_1$, nonlinear capacitor $C_1$  and parallel resistor  $R_p=Q/(L_1\omega _0)$, where $Q$  is the quality factor and $\omega _0$ the lowest frequency of the harmonic modal spectrum of the transmission line. There is a linear coupler  $C_2$ and $L_2$  between cells. This unit resonator has a soft nonlinearity because of increased capacitance at larger voltages. The voltage dependence of the MOS-FET capacitors was measured and 32 FETs were selected from more than 100 candidates. The experimentally determined saturable capacitance as a function of the DC bias voltage is shown in Fig.~\ref{fig:fig1}. The main difference between this transmission line and the one in Ref.~\cite{4} is that the current one has a somewhat narrower pass bandwidth, and no extra diodes, so that spatially sharp ILMs and an LSM locked to the driver can be studied even for such a small lattice. A cw driver excites the entire lattice uniformly to maintain either ILMs or LSMs in a large amplitude state. At each lattice cell $n$ the voltage $V_n$ is measured with an oscilloscope. Power is transferred via a small capacitor  $C_d$ to ensure a weak coupling between the driver and the circuit. The values and standard deviations (SD) of these electrical components are given in Table~\ref{tab:1}. The linear frequency band ranges from 287.30 kHz to 301.86 kHz. All coils are adjustable and were carefully tuned to the same value before starting the experiments.

\begin{table}
\caption{\label{tab:1}
Experimental electrical parameters and values used in the simulations. Those are approximately the same for experiment and simulations. The linear frequency pass band is from 287.30 kHz to 301.86 kHz.
}
\begin{tabular}{ccccc}
symbol & $C_d$(pF) & $L_1$($\mu $H) & $C_2$(pF) & $L_2$($\mu $H) \\
\hline
mean & 34.09 & 313.74 & 421 & 626.08 \\
(SD/mean)\% & 0.244 & 0.108 & 0.1 & 0.315 \\
simulation & 34.1 & 313.7 & 421 & 626 \\
\hline\hline
$Q$ & $k_1$(pF) & $k_2$(pF) & $k_3$(V) & $A_d$(V) \\
\hline
58.9 & 1613.33 & -669.33 & 2.58 & 10.5 \\
4.6 & 0.351 &  &  &  \\
50.0 & 1613 & -669 & 2.58 & 10.0 \\
\end{tabular}
\end{table}

\begin{figure}
\includegraphics{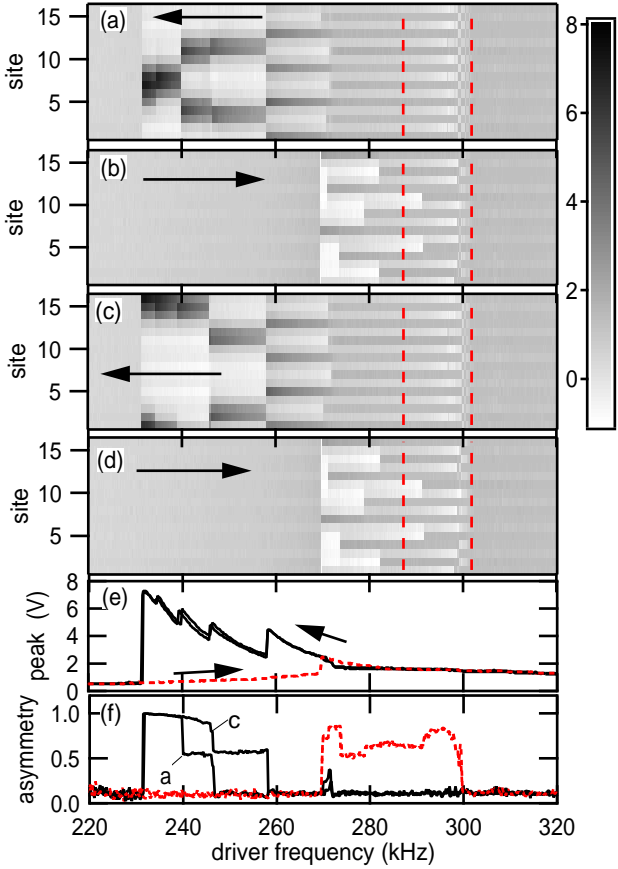}
\caption{\label{fig:fig2}Measured frequency dependence of driver locked ILMs and LSMs for the 16 element nonlinear cyclic transmission line. Four sweeps illustrate conversions. The nonlinear elements are saturable capacitors. Vertical dashed lines indicate the bottom (287.30 kHz) and top (301.86 kHz) of the small amplitude modal spectrum. Experimentally measured $Q=58.9$. Panels (a,c) Vibration vector at each frequency. Driver frequency starts above the modal spectrum. Arrow shows direction. Darker gray represents larger peak voltage. Nearly identical AR LSM patterns continuously changes into AR ILMs at about 270kHz. With decreasing frequency the ILM number decreases from 4 to 2 to 1. Further decrease of the driver frequency ends the stable ILM state. Panels (b,d) Increasing driver frequency from the low amplitude state (230 kHz). No ILM production but AR LSM appears suddenly at about 270 kHz.  These nearly identical AR LSMs are tuned through the normal mode spectrum and vanish at the top. (e) Sum of the lattice peak voltages as a function of the driver frequency for (a,c) down-scans (solid) and (b,d) up-scans (dashed). (f) Asymmetry (see Eq.~\ref{eq:1}) for both sweep directions versus frequency.}
\end{figure}

A measurement at a particular driving frequency consists of recording enough data synchronized to the driver that the maximum component  $V_n$ of the 16 element vector versus t can be identified. At a particular instant all cell voltages are identified relative to the driver and that is taken as the voltage vector for that particular driver frequency. For this driven system the initiation of the nonlinear spectral results depends on a modulational instability so that no two starting conditions are expected to give the same final results; hence, eight different runs were carried out. Two sets of such results are shown and analyzed in Fig.~\ref{fig:fig2}, where panels (a,b,c,d) present the single time voltage cell pattern as a function of the driver frequency F; the darker the gray the larger the voltage at that site. Since the nonlinear excitations are locked to the driver frequency they all have the same phase relative to the driver phase, and hence, most of the components of the voltage vector presented are positive. In panels (a,c) the frequency is swept down (see arrow) through the small amplitude modal spectral region identified by the two vertical dashed lines. Upon reaching the gap region the relatively fixed AR LSM pattern is transformed into ILMs. They decrease in number with decreasing driver frequency. A detailed measurement of the linewidth of the ILM shows initially it decreases with increasing ILM frequency shift from the modal spectrum but then increases because the nonlinearity of the capacitor saturates with increasing driving voltage. \cite{4, 27} When sweeping the driver frequency to larger values, as shown in panels (b,d), the dotted traces indicate that only the AR LSM is produced in the modal region. To present these results another way the voltage value at each cell in panels (a,c) or (b,d) is summed and plotted versus frequency in panel~\ref{fig:fig2}(e). The solid curve is for decreasing sweep frequency while the dotted one is for increasing frequency. Because of the complexity of the results shown in panels (a,b,c,d) a relevant question is how uniformly balanced is the AR energy pattern around the ring? To address this question the energy asymmetry, defined as 
\begin{equation}
asym = \frac{{\sum\limits_{n = 1,N/2} {\left| {\left| {V_n } \right|^2  - \left| {V_{n + N/2} } \right|^2 } \right|} }}{{\sum\limits_{n = 1,N/2} {\left( {\left| {V_n } \right|^2  + \left| {V_{n + N/2} } \right|^2 } \right)} }}
\label{eq:1}
\end{equation}
for N even, is plotted versus driving frequency. When the system is balanced, such as for a harmonic system, $asym=0$ while for the maximum imbalance of Eq.~\ref{eq:1}, $asym=1$. In panel (f) the solid curves are for the downward sweeps and the dashed curves are for the upward sweeps. As expected the largest asymmetry occurs when only one ILM remains. Less obvious is the asymmetry for both sweep directions observed for the AR LSM in the modal spectral region.

By repeating such experiments a number of times it was possible to determine the importance of the low concentration of impurities represented by the SD shown in Table~\ref{tab:1}. For the down scan cases we found that the LSM amplitude peaks in the modal region most often occur at odd sites rather than equally distributed at odd or even sites, depending on the experimental run. For the up scan cases, although panels (b,d) in Fig.~\ref{fig:fig2} seem to show a disorganized pattern, repeated runs such as these display very similar signatures demonstrating again that even a very low concentration of impurities plays a defining roll in the pattern outcome for either sweep direction.

\section{Comparison of simulations and experiment}
To compare with experiment the transmission line equation used in simulations is
\begin{eqnarray}
&&\frac{{d^2 q\left( {V_n } \right)}}{{dt^2 }} + \frac{{V_n }}{{L_1 }} + \frac{{L_1 \omega _0 }}{Q}\frac{{dV_n }}{{dt}} = C_d \frac{{d^2 }}{{dt^2 }}\left( {V_d  - V_n } \right) \nonumber \\ 
&&+ \left( {\frac{1}{{L_2 }} + C_2 \frac{{d^2 }}{{dt^2 }}} \right)\left( {V_{n + 1}  + V_{n - 1}  - 2V_n } \right) 
\label{eq:2}
\end{eqnarray}
where $q$ is the charge and $V_d$ is the driving voltage. The nonlinear capacitor is closely fitted by 

\begin{equation}
C\left( V \right) = k_1  + k_2 \exp \left( -\left( V/k_3 \right)^4 \right)
\label{eq:3}
\end{equation}

where $k_1$, $k_2$  and $k_3$  are constants. This fit is represented by the dotted curve in Fig.~\ref{fig:fig1}. The definition of capacitance $C=dq/dV$  is consistent with the $C(V)$ measurement where DC bias voltage is applied and capacitance is calculated from small amplitude AC current and voltage. The final dynamical equation for cell $n$ is
\\
\\
\begin{eqnarray}
&&\left( {2C_2  + C_d  + C(V_n)}  \right)\frac{{d^2 V_n }}{{dt^2 }}- C_2 \frac{{d^2 }}{{dt^2 }}\left( {V_{n + 1}  + V_{n - 1} } \right) \nonumber \\ 
&&=4\frac{{k_2 }}{{k_3 }}\left( {\frac{{V_n }}{{k_3 }}} \right)^3 \exp \left( { - \left( {\frac{{V_n }}{{k_3 }}} \right)^4 } \right)\left( {\frac{{dV_n }}{{dt}}} \right)^2  - \frac{V_n}{{L_1 }}  \\
&&+ \frac{1}{{L_2 }}\left( {V_{n + 1}  + V_{n - 1}  - 2V_n } \right) - \frac{L_1 \omega _0}{{Q }}\frac{{dV_n }}{{dt}} + C_d \frac{{d^2 V_d }}{{dt^2 }}  \nonumber 
\label{eq:4}
\end{eqnarray}
where the last two terms identify damping and driving at frequency  $\Omega=2\pi F$. For simulations the driver term becomes
\begin{equation}
C_d \frac{{d^2 V_d }}{{dt^2 }} = C_d \Omega ^2 A_d \cos \Omega t
\label{eq:5}
\end{equation}
The inductance and capacitance values, listed in Table~\ref{tab:1}, are approximately the same as for the experimental circuit. No impurities are introduced here.

\begin{figure}
\includegraphics{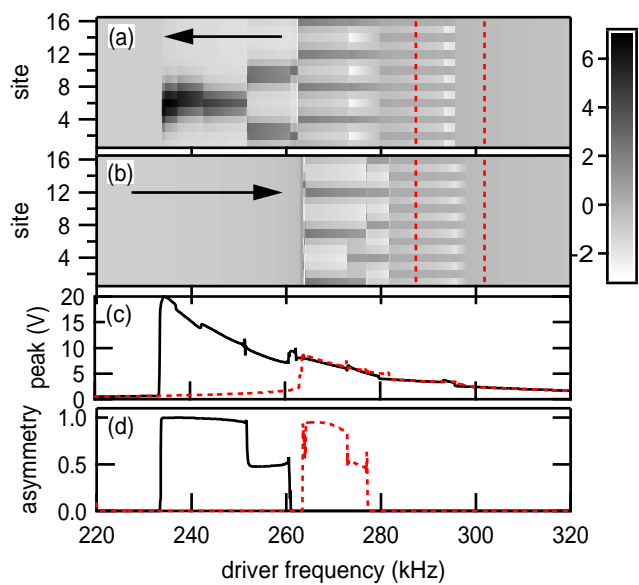}
\caption{\label{fig:fig3}Simulated maximum voltage of the 16 element lattice as a function of the driver frequency for $Q = 50$. Vertical dashed lines identify the top and bottom of the low amplitude modal spectrum. Arrows indicate scanning directions. (a) Simulation from above the top of the modal spectrum (301.86 kHz). The 8 peaked spatial standing wave first transforms into a 4 four peak LSM then into 4, 2 and 1 ILM of increasing width, respectively. (b) For the up scan a 3 peak LSM first appears at about 264 kHz. After some variability this converts into the 8-peaked LSM inside the modal spectrum. (c) Sum of the lattice peak voltages as a function of the driver frequency. Solid curve: down-scan from (a). Dashed curve: up-scan from (b). (d) Asymmetry for both sweep directions versus frequency.}
\end{figure}

The simulation method, including the addition of initial random noise, is described in Ref.~\cite{24} and in the supplement (suppl.pdf). Numerical results are presented in Fig.~\ref{fig:fig3} for the quality factor, $Q=50$, similar to the measured $Q$ of the experimental results in Fig.~\ref{fig:fig2}. This figure demonstrates that an AR LSM pattern connects directly to the AR ILM spectrum. Again the simulation direction starts from the arrow tails. In Fig.~\ref{fig:fig3}(a), four different patterns are observed as the frequency is swept down from the top of the modal spectrum. An LSM with 8 peaks  continuously changes into 2 ILMs below 260 kHz, the boundary between the AR ILM and the AR LSM. The evolution continues until only a single ILM remains. When the driver frequency is increased from below the ILM bifurcation point, see Fig.~\ref{fig:fig3}(b), a no-ILM state is observed until 260 kHz where a 3-peak LSM pattern suddenly appears and converts to an 8-peak LSM. Presented in panel (c) is the sum of the voltage value of each element of the vector at each driving frequency, including both the down scan (solid curve) and the up scan (dashed curve). The last panel (d) presents the asymmetry [Eq.~\ref{eq:1}] for these results for both sweep directions. 

\begin{figure}
\includegraphics{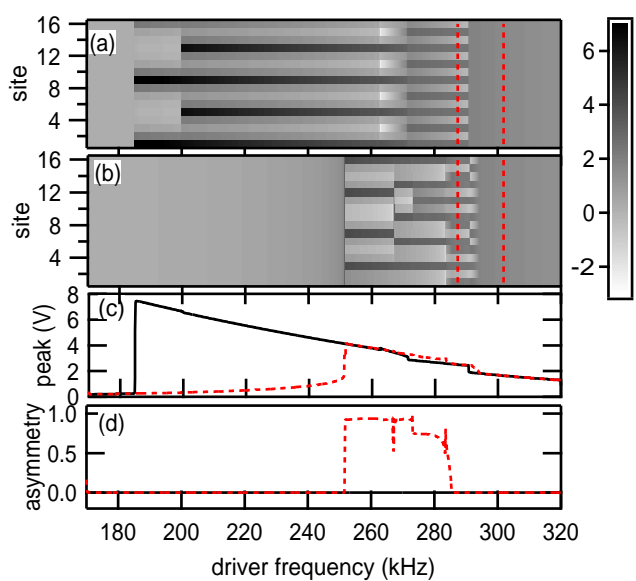}
\caption{\label{fig:fig4}Simulated frequency dependence of driver locked ILMs and LSMs for the 16 element nonlinear cyclic transmission line. The nonlinear elements are non-saturable capacitors. (a), (b) Sweep directions versus drive frequency. (c) Sum of the lattice peak voltages as a function of the driver frequency. Solid curve: down-scan from (a). Dashed curve: up-scan from (b). (d) Energy asymmetry results for both sweep directions versus frequency. Balanced ILMs and LSM patterns on the downward sweep and unbalanced LSM patterns on the upward sweep.}
\end{figure}

\begin{figure}
\includegraphics{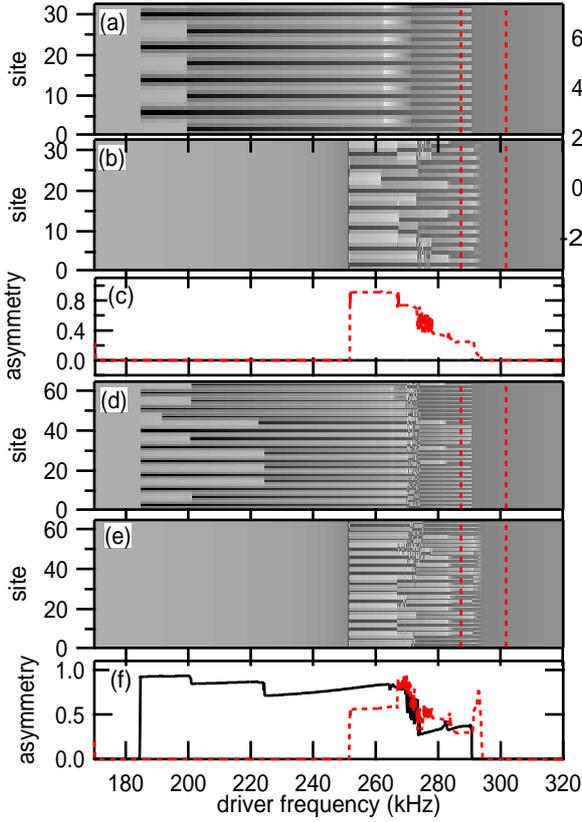}
\caption{\label{fig:fig5}Simulation results for a driven lattice with non-saturable capacitor for $N=32$ (a-c) and $N=64$ (d-f). Panels (a)-(b) show results for down scanning and up scanning cases, respectively. Panel (c) displays the energy asymmetry for down scanning (solid) and up scanning (dashed). Panels (d)-(e) present results for down scanning and up scanning cases, respectively. Panel (f) shows energy asymmetry for down scanning (solid) and up scanning (dashed). A down scan asymmetric pattern such as this appears in about 50\% of the runs.  }
\end{figure}

\begin{figure}
\includegraphics{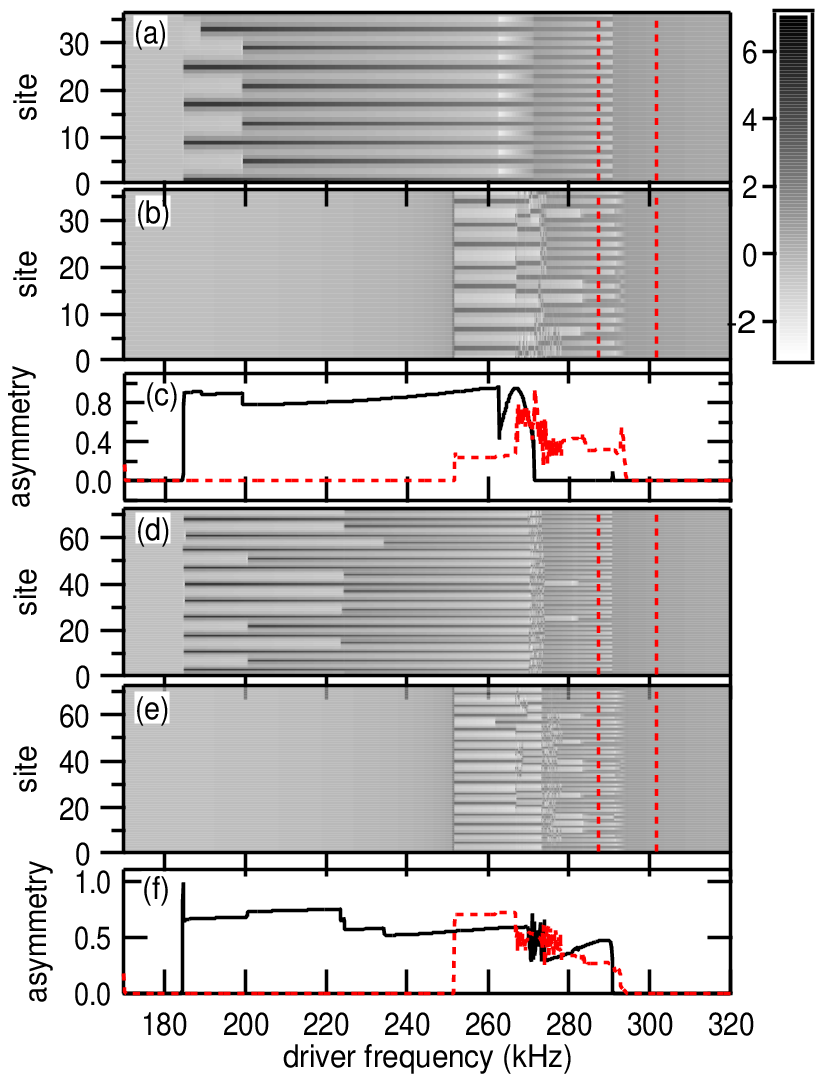}
\caption{\label{fig:fig6}Simulation results for a driven perfect lattice with non-saturable capacitor for $N=36$ (a-c) and $N=72$ (d-f). Panels (a)-(b) show results for down scanning and up scanning the driver frequency. Panel (c) presents energy asymmetry for down scanning (solid) and up scanning (dashed). Panels (d)-(e) display results for down scanning and up scanning, respectively. Panel (f) gives energy asymmetry for down scanning (solid) and up scanning (dashed). For both element numbers the down scan asymmetric patterns such as these appear in about 75\% of the runs.}
\end{figure}

\section{Simulations with a nonlinear capacitor for different cell numbers} 
So far the simulations have involved a saturable capacitor. To compare with the simulation results for a nonlinear capacitor without saturation the nonlinear expression in Eq.~\ref{eq:3} was replaced by  $C \sim V^2$, the dashed curve shown in Fig.~\ref{fig:fig1}. The simulated single shot voltage vector results of the 16 element lattice, as a function of the driver frequency for both sweep directions, are given in Fig.~\ref{fig:fig4}. Again the vertical dashed lines identify the top and bottom of the low amplitude modal spectrum. For the down sweep the LSM  converts to 4 ILMs. For the upsweep the 4 peak LSM transforms to an 8-peaked one. (The supplement (suppl.pdf) illustrates how the boundary between an LSM and equally spaced ILMs can be readily identified in simulations.) Panel (c) illustrates that the sum of the voltage components is fairly smooth in both sweep directions in contrast to that found in Fig.~\ref{fig:fig3} while panel (d) indicates no asymmetry for the downwards sweep 100\% of the runs. The LSM asymmetry now only appears in the up sweep direction both outside and inside the modal spectral region.

Given the small lattice size and the condition that an integer number of wavelengths of the LSM pattern and an integral number of ILMs must fit the cyclic boundary conditions it is not too surprising that Fig.~\ref{fig:fig3} and Fig.~\ref{fig:fig4} look different even though there are no impurities. Since the peaks for the saturable capacitor case broaden with increasing voltage the number of features may jump and hence the possibility of energy asymmetry becomes more likely. Presented with the relative simplicity of the patterns in Fig.~\ref{fig:fig4} for 16 elements it is worthwhile to examine the patterns as the cell number in the perfect lattice is increased.

Figure~\ref{fig:fig5} presents the simulation results for the number of elements doubled (a-c) and then doubled again (d-f). The down sweep and upsweep pattern for $N=32$ are very similar to that observed for the 16 element case in that there is no energy asymmetry in the down sweep but LSM asymmetry in the up sweep over the frequency region for 100\% of the runs. For the $N=64$ element case the lattice is now large enough that both ILM and LSM variations are evident. The asymmetry for the down sweep recorded in panel (f) occurs in about 50\% of the runs. For the up sweep cases, where the LSM excitation pattern suddenly appears randomness is more evident and this asymmetry appears in 100\% of the runs.

Changing the number of elements from $N=32$ to $N=36$ profoundly influences the asymmetry for the down sweep as shown in Fig.~\ref{fig:fig6}(a-c). At some point in this sweep the number of ILMs becomes 9 and that result ensures energy unbalance around the ring. Sweeps such as this occur for about 75\% of the runs. For $N=72$ the resulting pattern is qualitatively similar to that for $N=64$ and now asymmetry such as this appears in about 75\% of the downward sweeps and 100\% of the upward sweeps. It should be remembered that if a low concentration of random impurities would be included then the patterns would tend to become pinned and energy asymmetry would become more likely for all of these examples.

\section{Discussion and Conclusions}
This soft nonlinear lattice energetically prefers site-peaked ILM patterns of the S-T type \cite{7} and the most stable patterns are only equal-spaced. For the $N=16$ lattice, possible $k$-values are only $k =  \pm \pi $ (8 peaks),  $k =  \pm \pi /2$(4 peaks), and  $k =  \pm \pi /4$(2 peaks). In this $N=16$ element lattice, the LSM switches to ILMs at the middle of the 4 peak-frequency region. For the up-scanning cases, the LSM pattern is not completely equally spaced, but the peaks are all lattice site-centered. This occurs because the transition begins at the direct excitation of $k=0$ normal mode fed by the driver, and the system suddenly acquires enough amplitude to generate an LSM. Thus, the system transitions to one of the meta-stable states, which may not consist of equally spaced peaks; the larger the lattice the larger the number of available metastable states.

A driven nonlinear transmission line, with periodic boundary conditions and elements that contain a saturable nonlinear capacitor, has been used to generate AR ILMs outside of the modal spectrum and an LSM within it. In the experiments a very small impurity concentration was observed to influence the position of the LSM pattern when driving the frequency in either direction across the modal spectrum. The most dramatic feature is that by simply changing the driver frequency the spectrum can evolve continuously from a LSM pattern distributed around the ring to multiple ILMs localized on a few sites. This can occur either with or without impurities. The resultant  energy distribution, for either the experimental saturable capacitor or the perfect simulated unsaturable nonlinear capacitor, can either be balanced or unbalanced. Simulations show that unbalanced energy in the perfect nonlinear system becomes more likely as the number of elements is increased, as the $Q$ is increased or as the modal bandwidth is decreased. 

Given that the experimental results presented here stem from a general nonlinear model system applications of these findings may be expected in other fields. Rotational machinery, such as a gas turbine with nonlinear periodic blades, represents one particularly relevant example. The vibrational spectra of these complex systems, where ``engine order'' excitation can excite modes of multiple branches, have been studied for many decades\cite{28,29,30,31,32,33,34}. The observed fluttering of rotating gas turbine blades is very sensitive to mistuning and such mistuning can be produced by impurity or defect-induced local vibrational modes for the linear dynamical range. The observed high cycle blade fatigue has motivated multiple theoretical and experimental studies of nonlinear blades, of the contribution of interblade frictional devices as well as of blade root friction dampers to control vibration and yet frequently ``rogue blades'' crack, break or are expelled. 

From the work presented here it is clear that even a driven turbine bladed disc consisting of homogeneous nonlinear blades of exactly the same geometry and composition and hence ``tuned'' in the classical sense, can, given the right driving and friction damping conditions, generate large localized mechanical vibrations outside and LSM patterns inside the modal spectrum. Include a low concentration of geometrical impurities produced by machine tolerances and vibrational energy imbalance can occur both inside and outside a modal spectrum. For a demonstration consider the experimental results shown in Fig.~\ref{fig:fig2}. Driving a bladed disc with soft nonlinearity and large vibrational damping up in frequency toward its lowest modal spectrum will produce no ILMs but at a certain frequency a ILM/LSM pattern can develop around the array and be autoresonantly maintained as the driver frequency increases. Next if the driving frequency is now decreased through the same spectrum then the LSM converts to ILMs. An energy asymmetry pattern is very likely for both of these situations. The experimental hallmark is, given the right driving conditions, a rotational driven nonlinear bladed disc containing damping would remain in balance with increasing driving frequency up to the lowest modal spectrum but then become unbalanced at still higher frequencies. With decreasing driving frequency the same nearly perfect blade array already unbalanced in the modal spectrum region would become more unbalanced via ILM production. This vibration signature during cycling, which is very different from that produced by geometric impurity modes, would be the case for soft nonlinearity. For a bladed disc containing hard nonlinear periodic blades, by symmetry, one can expect that the same sort of driven vibrational energy asymmetry for the rotating bladed disc will again appear, but now mainly due to ILMs generated above its modal spectrum.


\acknowledgments
MS is supported by JSPS Kakenhi Grant number JP16K13716. AJS acknowledges the hospitality of the Department of Physics and Astronomy, Denver University, where some of this work was started.

\providecommand{\noopsort}[1]{}\providecommand{\singleletter}[1]{#1}%

\end{document}